\documentclass[prx,twocolumn,notitlepage,aps,showpacs,longbibliography,amsmath,amstex,amssymb,mathfonts]{revtex4-1}

\pdfoutput=1

\usepackage[pdftex]{graphicx}

\usepackage{amsmath}
\usepackage{amssymb}
\usepackage{mathtools}

\usepackage[protrusion=true,expansion=true]{microtype}

\usepackage{hyperref}
\hypersetup{
	pdftoolbar=true,        		 
	pdfmenubar=true,     		 
	pdffitwindow=false,     	 
	pdfstartview={FitH},    	 
	pdftitle={String order parameters for 1d Floquet Symmetry Protected Topological Phases},   
	pdfauthor={Kumar, Dumitrescu, Potter},     		 
	pdfsubject={},   			 
	pdfcreator={},   	        	   	 
	pdfproducer={}, 		
	pdfnewwindow=true,      	
	colorlinks=true,       		
	linkcolor=black,          	
	citecolor=blue,        		
	filecolor=magenta,    		
	urlcolor=blue          		
}

\usepackage{color}
\usepackage{array}
\usepackage{bbm}
\usepackage{hyperref}
\usepackage{color}
\usepackage{array}
\usepackage{cancel} 
\usepackage{ulem} 

\usepackage{booktabs}
\newcolumntype{C}[1]{>{\centering\arraybackslash}m{#1}}
\AtBeginDocument{
\heavyrulewidth=.08em
\lightrulewidth=.05em
\cmidrulewidth=.03em
\belowrulesep=.65ex
\belowbottomsep=0pt
\aboverulesep=.4ex
\abovetopsep=0pt
\cmidrulesep=\doublerulesep
\cmidrulekern=.5em
\defaultaddspace=.5em
}
\usepackage{booktabs}
\newcolumntype{R}[1]{>{\raggedleft\arraybackslash}p{#1}}
\AtBeginDocument{
\heavyrulewidth=.08em
\lightrulewidth=.05em
\cmidrulewidth=.03em
\belowrulesep=.65ex
\belowbottomsep=0pt
\aboverulesep=.4ex
\abovetopsep=0pt
\cmidrulesep=\doublerulesep
\cmidrulekern=.5em
\defaultaddspace=.5em
}

\newcommand{\<}{\langle}

\newcommand{\up}{\uparrow}
\newcommand{\down}{\downarrow}
\renewcommand{\>}{\rangle}
\renewcommand{\(}{\left(}
\renewcommand{\)}{\right)}
\renewcommand{\[}{\left[}
\renewcommand{\]}{\right]}

\newcommand{\Z}{\mathbb{Z}}

\newcommand{\red}[1]{\textcolor{red}{#1}}

\newcommand{\subheader}[1]{\vspace{4pt}\noindent{\bf #1 -- }}

\begin{document}

\title{String order parameters for 1d Floquet Symmetry Protected Topological Phases}

\author{Ajesh Kumar}
\affiliation{Department of Physics, University of Texas at Austin, Austin, Texas 78712, USA}
\author{Philipp T.~Dumitrescu}
\email{philippd@utexas.edu}
\affiliation{Department of Physics, University of Texas at Austin, Austin, Texas 78712, USA}
\author{Andrew C.~Potter}
\affiliation{Department of Physics, University of Texas at Austin, Austin, Texas 78712, USA}

\begin{abstract}
Floquet symmetry protected topological (FSPT) phases are non-equilibrium topological phases enabled by time-periodic driving. FSPT phases of 1d chains of bosons, spins, or qubits host dynamically protected edge states that can store quantum information without decoherence, making them promising for use as quantum memories. While FSPT order cannot be detected by any local measurement, here we construct non-local string order parameters that directly measure general 1d FSPT order. We propose a superconducting-qubit array based realization of the simplest Ising-FSPT, which can be implemented with existing quantum computing hardware. We devise an interferometric scheme to directly measure the non-local string order using only simple one- and two- qubit operations and single-qubit measurements.
\end{abstract}

\maketitle

\noindent 
Time-periodic (Floquet) driving enables fundamentally new symmetry-protected topological (SPT) phases of matter with dynamical properties that could not occur in thermal equilibrium~\cite{kitagawa2010topological,jiang2011majorana,rudner2013anomalous,von2016phaseI,else2016classification,potter2016topological,roy2016abelian,po2016chiral,roy2016periodic,harper2016stability}. Floquet SPT (FSPT) phases in $1d$ chains of interacting spins or qubits exhibit protected edge modes that undergo a repeating sequence of topologically protected spin-echoes that dynamically decouples them from bulk sources of decoherence~\cite{potter2016topological,potirniche2016floquet}.
In MBL settings these edges can store quantum information in a topologically protected manner, without the need to cool the system near its ground-state~\cite{bahri2015localization,potter2016topological,potirniche2016floquet}. 
Practically, $1d$ FSPT phases actually have less stringent symmetry requirements than their equilibrium counterparts~\cite{potter2016topological,potirniche2016floquet}. This feature allows them to be implemented with simple, realistic two-particle interactions, and paves the way for their realization in a variety of experimental setups including trapped ions, Rydberg atoms~\cite{potirniche2016floquet}, and superconducting qubits.

The topological structure of $1d$ FSPTs can be formally understood by mapping the time-periodicity of the drive onto an effective additional discrete time-translation symmetry~\cite{else2016classification,potter2016topological}. This enables a systematic classification and characterization of their topological properties via group cohomology methods~\cite{von2016phaseI,else2016classification,potter2016topological} or via topological features in the entanglement dynamics~\cite{potter2016topological,potirniche2016floquet}. 

Despite the theoretical progress in understanding these non-equilibrium topological phases, a viable method to directly measure the dynamical topological invariants of FSPT phases remains elusive.
In fact, the global topological properties of FSPT phases cannot be revealed by any local measurements (though it may be indirectly inferred by the presence of robust edge states). In this paper, we construct a non-local dynamical string order parameter to directly measure the $1d$ FSPT order, by extending a related construction for static SPTs to the dynamical realm \cite{nijs1989preroughening-staticstring, oshikawa1992hidden-z-2-staticstring, kennedy1992hidden-z2-staticstring}. 

After developing the general theory of such string-order-parameters for FSPT phases, we specialize to the simplest example of an FSPT protected by an Ising ($\Z_2$), spin-flip symmetry, and demonstrate that long-range string order can be used as a numerical diagnostic for this FSPT order. We then propose a simple protocol to realize this Ising-FSPT phase in transmon superconducting qubit arrays. Despite the complicated and non-local form of the string order parameters, we outline a practical interferometric scheme to measure them in recently developed two-leg ladders of superconducting qubits.

\subheader{FSPTs as charge pumps}
To formally establish the existence of dynamical string order in $1d$ FSPT phases, we focus on an FSPT phase with $\Z_n$ symmetry~\footnote{This readily extends to generic unitary symmetry groups which, for MBL systems, are necessarily finite Abelian groups, which decompose as: $\Z_{n_1}\times\Z_{n_2}\times \cdots \times \Z_{n_N}$~\cite{potter2016topological}.}. 
The generator $g$ of the $\Z_n$ symmetry has eigenvalues $e^{2\pi i q/n}$, where $q\in \{0,1,\dots,n-1\}$ is the charge associated with the symmetry.  We can generally view the phase as a chain of lattice sites, with an on-site representation of the symmetry. Define operators $\hat{q}^\pm_x$ that raise or lower the symmetry charge on site $x$, so that $g^\dagger \hat{q}^\pm g = e^{\pm2\pi iq/n}\hat{q}^\pm$. 

\begin{figure}[b]
\centering
\includegraphics[width=0.8\columnwidth]{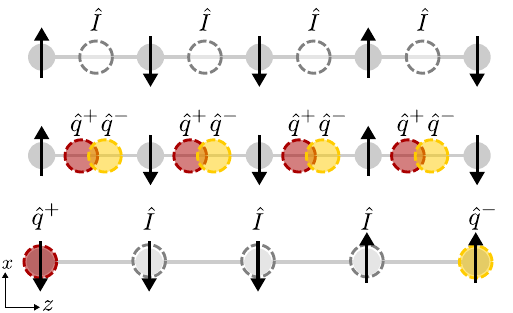}
\caption{{\bf Charge Pump Schematic. -- }  The FSPT evolution during one period is a topological charge pump. The topological aspect of $U(T)$ can be interpreted as locally nucleating dipoles of symmetry charge $\hat{q}^+\hat{q}^-$. These dipoles annihilate with neighboring dipoles and transfer a net symmetry charge across the system. Non-topological evolution gives additional trivial MBL dynamics which is not shown.
\label{fig:dipolecartoon} }
\end{figure}

Our construction derives from an intuitive picture of the formal FSPT classification as a charge pump~\cite{potter2016topological}.
In the FSPT phase, the Floquet operator $U(T)$ acts as a ``quantum Archimedes' screw'', pumping a quantized symmetry charge across the system and onto the ends of the chain. 
This global charge pumping is mediated by local charge transfer between nearby sites, as illustrated in Fig.~\ref{fig:dipolecartoon}. While there is no net accumulation of charge in the bulk after a full period, a charge $q$ is removed from the left edge of the chain and added to the right edge. 
Since the symmetry charges take discrete values, the amount of charge pumped cannot be continuously changed by any symmetric, T-periodic perturbation: it is a topological invariant of the FSPT phase.

The periodic pumping cycles the charge of the edge through all possible charges, thereby precisely canceling out any net coupling between the edge and the bulk. This results in topologically protected edge states. This dynamical decoupling is analogous to a spin-echo protocol on a single spin. Crucially, however, the pulse in a spin-echo protocol needs to be precisely tuned to achieve exact refocusing; the FSPT achieves perfect decoupling over a range of driving parameters due to the topological protection.

\subheader{Dynamical string order}
The charge pumping picture suggests a simple way to construct a non-local string order parameter that detects the FSPT order even far from the boundaries.
The idea is to induce an artificial edge, by truncating the Floquet evolution to a finite interval, and then to measure the accumulated charge. Specifically, suppose the FSPT is produced by a local time-dependent Hamiltonian $H(t) = \sum_x h_x(t)$, where each term $h_{x}(t)$ acts only within a finite-region around site $x$. We define the restriction of the evolution to an interval $[i,j]$ as $U_{[i,j]}(T) = \mathcal{T}\exp\{{-i\int_0^T dt\sum_{x\in [i,j]} h_x(t)}\}$. 

Just as the full Floquet operator $U$ pumps a symmetry charge across the entire system in the FSPT phase, $U_{[i,j]}$ pumps one across the interval $[i,j]$. We need to compensate this charge accumulation by acting with $\hat{q}_i^-, \hat{q}_j^+$ to obtain the actual non-local order parameter~\footnote{The ordering of the $\hat{q}^\pm$ operators relative to the $U_{[i,j]}$ in this definition is inessential as they are related by unitary transforms and encode equivalent information, here we have chosen a convention to agree with the equilibrium SPT string orders.}:
\begin{align}
S_{[i,j]} = \hat{q}_{i}^{-} U_{[i,j]} \hat{q}_{j}^{+}.
\label{eq:fstring}
\end{align}

When acting on an MBL eigenstate, $U_{[i,j]}$ preserves the state of local integrals of motion (LIOM)~\cite{vosk2013many,serbyn2013local,huse2013localization} deep inside the interval up to exponential accuracy in $\sim |i-j|/\xi$, yielding only an overall phase. Apart from this phase, $U_{[i,j]}$ acts non-trivially only near the edge of the interval, where it deposits or removes charge. 
Since the charge pumped by $U_{[i,j]}$ generically spreads over a region of characteristic size the localization length $\xi$, it is not precisely cancelled by the site-local operators $\hat{q}_i^-, \hat{q}_j^+$. Nevertheless, for finite $\xi$, these operators remove the pumped charge with non-zero fidelity, resulting in a non-zero expectation value of $S_{[i,j]}$, even for arbitrarily long intervals: $\lim_{{|i-j|}\rightarrow \infty}|\<n|S_{[i,j]}|n\>| > 0$. 
In contrast, for a non-topological Floquet drive, $U_{[i,j]}$ does not pump any charge and $S_{[i,j]}$ is equivalent to measuring the correlations in the symmetry breaking operators, which fall off exponentially with distance
$\lim_{{|i-j|}\rightarrow \infty}|\<n|S_{[i,j]}|n\>|\sim |\<n|\hat{q}_i^- \hat{q}_j^+|n\>| \sim e^{-|i-j|/\xi}$~
\footnote{For this reason, $S_{[i,j]}$ will also show long-range order for a time-translation invariant, but $\Z_n$ symmetry breaking system. Such conventional order can easily be distinguished from the FSPT by a concurrent measurement of the local Landau order parameter.}. 
The string operator $S_{[i,j]}$ therefore acs as a non-local order parameter distinguishing the FSPT and trivial phases.

A minor complication is that the string order exhibits glassy behavior --
while the amplitude of $\<n|S_{[i,j]}|n\>$ remains non-zero for a FSPT state, its phase will randomly vary with $i$ and $j$, in a state-dependent fashion. This is analogous to magnetic order behavior of a spin-glass. To avoid that the eigenstate average vanishes to this trivial phase variation, we consider an Edwards-Anderson type order parameter
\begin{align}
C_S(i,j) = \frac{1}{|\mathcal{H}|}\sum_n |\<n|S_{[i,j]}|n\>|^2.
\label{eq:stringEA}
\end{align}
Here $|\mathcal{H}|$ is the total number of states of the system. 

While convenient for simulations, \eqref{eq:stringEA} is unsuitable for experiments since it is not possible to directly prepare an exact Floquet eigenstate $\vert n\>$. Instead, one must look for temporal string correlations frozen into the long time dynamics by measuring
\begin{align}
C^{(d)}_S(i,j;t) = \<\psi_0| S_{[i,j]}^{\dagger}(t)S_{[i,j]}(0)|\psi_0\>
\label{eq:dyn_string}
\end{align}
in a quantum quench from an arbitrary initial state $|\psi_0\>$. On averaging over long time-intervals and over different initial states $|\psi_0\>$, $C^{(d)}_S$ exhibits a transient decay from $1$ to saturate at a non-zero value $\approx C_S$ at long times.

\subheader{Dual perspective}
A complementary perspective on the string order parameter comes from a dual description of $1d$ SPT states as condensates of charged domain walls (DWs)~\cite{chen2014symmetry}. For example, an equilibrium $1d$ SPT state with symmetry group $\Z_n\times \Z_m$, can be constructed from a $\Z_n$-symmetry breaking state by condensing DWs of the $\Z_n$ order that are bound to charges of the $\Z_m$ symmetry (see Appendix~\ref{app:dual}). Kramers-Wannier duality then maps the non-local $\Z_m$-charged DW condensate into a local $\Z_n\times \Z_m$-breaking dual-magnetic order, whose Landau order parameter is just the dual of the string-order parameter.

A related construction applies for FSPTs, which can be mapped onto $1d$ equilibrium SPTs with an extra time-translation symmetry~\cite{else2016classification,potter2016topological,yao2017discrete}. We can then obtain an FSPT state from DW condensation in a state that spontaneously breaks this dynamical time-translation symmetry -- a discrete time-crystal~\cite{khemani2015phase,else2016floquet,yao2017discrete,zhang2017observation,choi2017observation}. 
A discrete time-crystal exhibits persistent $nT$-periodic oscillations, of which there are $n$ distinct oscillating patterns that are related by evolving for one period $U(T)$, analogous to the $n$ different magnetic domains of a $\Z_n$ magnet which are related by applying the symmetry generator. 
By analogy to the equilibrium construction above, an FSPT state can be obtained from a time-crystal by condensing domain walls in the time-crystalline order bound to symmetry charges of $G$. 

To relate this dual description to the string order parameter, note that, when acting on a time-crystal state, the $U_{[i,j]}$ part of the string order parameter just shifts the phase of the time-crystal domain in $[i,j]$ by one cycle, creating a time-crystal DW (anti-DW) at the boundaries of the interval $[i,j]$. Concurrently, the edge-charge operators, $q^{+}_i$, ($\hat{q}^-_j$) add (remove) symmetry-charge $q$ to the locations of the time-crystal DW (anti-DW) respectively. Therefore, long range order in $S_I$ corresponds exactly to long-range order in charged-DW creation and annihilation operators, i.e. the string order parameter detects the charged-DW condensate.

\subheader{Superconducting Qubit implementation}
Having deduced a non-local string order parameter that detects a general FSPT invariant, we now propose a scheme to experimentally realize the simplest FSPT phase with a $\Z_2$ (Ising) symmetry and measure its non-local string order using arrays of superconducting transmon qubits. Dramatic progress in superconducting qubit technology has led to recent demonstrations of linear~\cite{kelly2015state} and two-leg ladder arrays~\cite{IBMPress} with 10's of qubits. The long coherence times and programmable single- and two-qubit gates in these systems provide an ideal platform for exploring non-equilibrium Floquet phases. Conversely, MBL and FSPT physics are naturally advantageous for these systems, as they can be achieved in non-equilibrium quenches from an arbitrary initial state and do not require the difficult preparation of correlated many-body eigenstates.

\begin{figure}[t]
\centering
\includegraphics[width=\columnwidth]{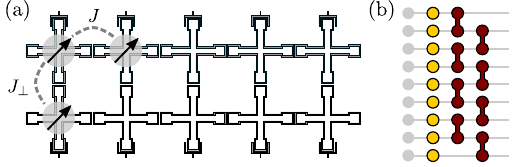}
\caption{{\bf Experimental Protocol. -- } (a) Schematic of a two-leg ladder of superconducting transmon qubits. After implementing the FSPT phase in each one-dimensional array, we can use the inter-leg couplings $J_\perp$ to measure the non-local string order parameter. (b) Three step protocol \eqref{eq:3stepdrive} realizing 1$d$ $\Z_2$ FSPT phase: random on-site gates (gold), followed by two sets of pair-interactions between the qubits (red).
\label{fig:setupseq} }
\end{figure}

One obstacle to obtaining a $1d$ FSPT realization in these systems is that the two-qubit interactions naturally take the form $H_J(t) = -\sum_i J_i(t)(X_iX_{i+1}+Y_iY_{i+1})$, whose $U(1)$ symmetry does not, by itself, support an FSPT phase. In Appendix \ref{sec:u1toising}, we show that a sequence of staggered two-qubit operations and single-qubit rotations can be used to generate effective Ising interactions $H_\text{Ising} = \sum_i J_i(t)X_i X_{i+1}$. These enable a variety of non-equilibrium phases including MBL spin-glasses~\cite{huse2013localization} and discrete time-crystals~\cite{khemani2015phase,else2016floquet,yao2017discrete,zhang2017observation,choi2017observation}. 

\begin{figure}[t]
\centering
\includegraphics[width=\columnwidth]{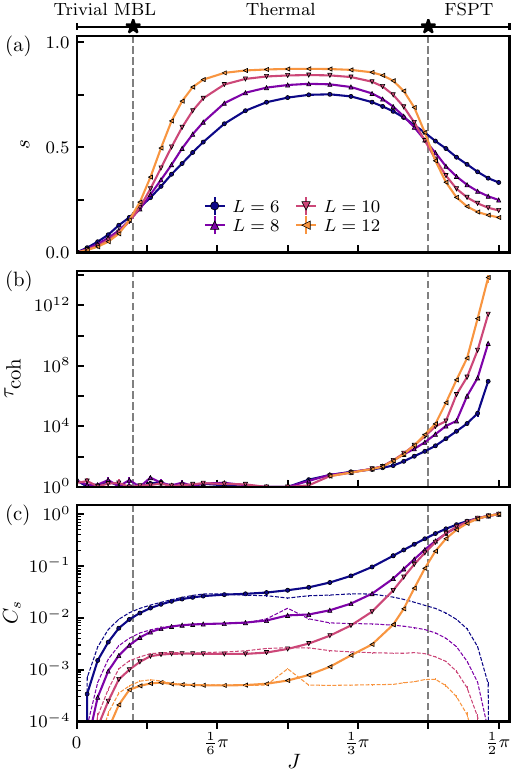}
\caption{{\bf Phases of the three-step drive. -- } Eigenstate observables as a function of the two-qubit interaction $J$, for different system sizes $L$, and averaged over $5\cdot 10^2-10^4$ disorder realizations. Vertical dashed lines indicate approximate phase boundaries from entanglement. (a,b) have open boundaries, and (c) has periodic boundary conditions.  (a) Normalized bipartite entanglement entropy averaged over $2^5$ eigenstates distinguishes MBL and thermal phases. (b) Fit coherence time $\tau_{\mathrm{coh}}$ of the edge spin diverges exponentially with $L$ in the FSPT phase, which becomes infinite at $J=\pi / 2$. Error estimates are obtained by bootstrap resampling.
(c) Non-local string order parameter from \eqref{eq:stringEA} on a closed system, which directly detects FSPT order despite the absence of edge-states. %
\label{fig:saticobs} }
\end{figure}

For the purpose of realizing a $\Z_2$ FSPT, this protocol can be compressed into a simpler one with only three steps  per Floquet period~(Fig.~\ref{fig:setupseq}b):
\begin{align}
H(t) = \frac{3}{T}\begin{cases}
\sum_i h_i X_i & \phantom{\tfrac{1}{3} T}\makebox[0pt][r]{0}\leq t< \tfrac{1}{3} T \\
\sum_{i~\textrm{odd}} J\left(X_iX_{i+1}+Y_iY_{i+1}\right)   &  \tfrac{1}{3} T \leq t < \tfrac{2}{3} T \\
\sum_{i~\textrm{even}} J\left(X_iX_{i+1}+Y_iY_{i+1}\right) &  \tfrac{2}{3} T\leq t < T 
\end{cases}
\label{eq:3stepdrive}
\end{align}
The random fields $h_i\in [-h,h]$ both induce MBL and break down the $U(1)$ symmetry of the two-qubit gates to a $\Z_2$ Ising symmetry generated by $g= \prod_i X_i$. The last two sequences of pulses lead to symmetry charge pumping and edge states when the system is in the FSPT phase. On  an open chain of size $L$ at parameter $J=\pi/2$, the time evolution is simply $U(T) = Z_1Z_Le^{-i\sum_i h_iX_i}$. This realizes the zero correlation-length limit of the FSPT. The operators $Z$ act on the left and right ends of the chain, flipping the edge spins in the $x$ basis. The flipping exactly corresponds to adding a charge $Z_i = \hat{q}^+_i$ or removing a charge $Z_i = \hat{q}^-_i$ -- these two operations are equivalent for a $\Z_2$ symmetry. Over the course of two periods the edge-spin undergoes a spin-echo process that decouples it from the bulk motion which is $T$-periodic.

We numerically map out the phase diagram with varying $J$, simulating arrays of up to 12 qubits. Disorder is fixed to its maximal value $h=\pi$ and $T=1$ is the unit of time. We observe three distinct phases: for small $J$, the system is a trivial MBL paramagnet, for $J$ close to $\pi/2$ the system is an MBL FSPT, and for intermediate $J$, the system thermalizes.

To distinguish the MBL and thermal regions, we compute the half-system entanglement $S\(\frac L2\)$, averaged over Floquet eigenstates and disorder realizations. Normalizing by the leading order infinite-temperature thermal value gives $s(L) = S\(\frac L2\)/\(\frac L2\log 2\)$.  Upon increasing $L$, $s$ exhibits two finite size crossing at moderate $J$ which separates the MBL phases ($s \sim 1 / L$) and thermal phase ($s \sim 1$); see Fig.~\ref{fig:saticobs}a.

We diagnose the FSPT order in two ways. First, via the coherence time $\tau_\text{coh}$ of the edge spins in an open chain, by fitting the disorder averaged spin-correlations $\<Z_1(t)Z_1(0)\>$ to an exponential form $\sim e^{-t/\tau_\text{coh}}$ (Fig.~\ref{fig:saticobs}b). Near $J=\pi/2$, $\tau_\text{coh}$ increases rapidly, and also diverges with system size $L$ as expected for topologically protected edge-states. Second, we compute the eigenstate averaged string order parameter, $C_S(1,L/2 + 1)$ in a periodic chain (Fig.~\ref{fig:saticobs}c). We observe finite size behavior consistent with long-range spin-glass string-order. At small and intermediate $J$ it decreases exponentially with system size $L$ as expected for disordered systems. At a critical $J_c$, the order onsets and grows to $1$ at $J = \pi / 2$; in the ordering region $C_S(1,L/2 + 1)$ saturates to a non-zero constant with increasing $L$. 

\subheader{Measuring the non-local order parameter}
An obvious challenge to observing the string order parameter is its complicated and non-local form. This obstacle can be surmounted by making two copies of the system, and performing a quantum interferometric sequence that maps the string order parameter to a simple set of single-qubit measurements in the computational basis. 

\begin{figure}[t!]
\centering
\includegraphics[width=\columnwidth]{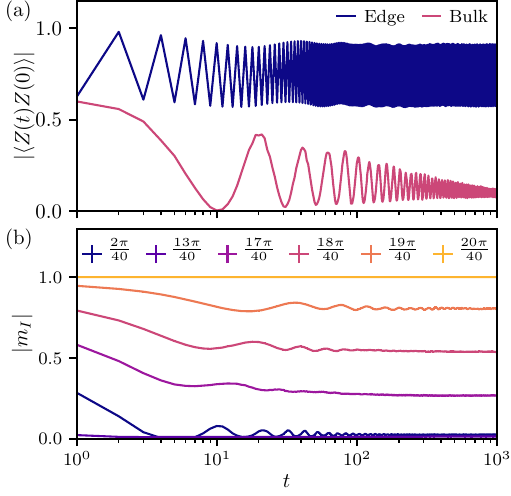}
\caption{{\bf Correlation Functions detecting FSPT order at $L=9$ -- }  (a) Correlation functions in the FSPT phase with  $J = 19 \pi / 40$. The edge spins oscillates coherently up to infinite times; a bulk spin loses coherence. To show the envelope of the oscillations, we take the absolute value before averaging over disorder realizations. (b) Time dependance of the string order parameter shows a decay and saturation to $\sim C_s$ in the FSPT. The decay of the observable in the thermal phase ($J = 13 \pi / 40$) is much faster than in the trivial MBL phase ($J=2\pi/40$).
\label{fig:timesweeps} 
}
\end{figure}

Our scheme realizes copies of the FSPT phase in a $L\times 2$ array of qubits that has local transmon coupling between the two chains (Fig.~\ref{fig:setupseq}a).  Such two-leg ladder devices with $L=8$ have recently been experimentally realized by the IBM group~\cite{IBMPress}. The idea is to initialize the two chains into the same state, $|\psi\>$, so that the overall system is in state $|\Psi\> = |\psi\>\otimes|\psi\>$. Next, we evolve the first row of qubits with time-evolution $V_n = S_{[i,j]}U(nT)$, and the second with $V'_n=U(nT)S_{[i,j]}$. Both $V$ and $V'$ can be readily implemented by selectively applying the sequence \eqref{eq:3stepdrive} to the spins inside the interval ${[i,j]}$, or to the entire chain in the appropriate order, resulting in $|\Psi'\> = (V_n|\psi\>)\otimes(V'_n|\psi\>)$. Next, suppose that one is able to measure the SWAP operator $\Sigma$, which exchanges the states of the two systems, defined by $\Sigma|\psi_1\>\otimes|\psi_2\> = |\psi_2\>\otimes|\psi_1\>$. For the above choice of $V$ and $V'$, the SWAP measurement simply produces the dynamical string-correlations:
\begin{align}
\<\Psi'|\Sigma|\Psi'\> &= |\<\psi|U^\dagger(nT)S_{[i,j]}U(nT) S_{[i,j]} |\psi\>|^2 \nonumber\\
&= |C^{(d)}_S(i,j;nT)|^2
\end{align}
In pioneering experiments~\cite{islam2015measuring}, a protocol to measure entanglement via SWAP operators~\cite{horodecki2002method} was experimentally implemented in bosonic optical lattice systems. We can adapt this boson-based scheme for use in qubit arrays as follows (see also \cite{[{}][{}]garcia-ripoll2004implementation}).

Consider the pair of qubits in a given column. Viewing each qubit as an effective spin-1/2, we can describe the four-possible states of this pair by the singlet and triplet states, and consider the action of SWAP in this column-pair basis. Only the column singlets transform under a SWAP operation, acquiring a $(-1)$ phase. Therefore, the overall SWAP eigenvalue is equal to $(-1)^{N_s}$ where $N_s$ is the number of column-singlets. To facilitate measurement of $N_s$ we can first perform the unitary operation: 
\begin{align}
W =& e^{-i\frac{\pi}{8}\sum_{j=1}^L\(X_{1,j}X_{2,j}+Y_{1,j}Y_{2,j}\)}e^{i \frac{\pi}{8}\sum_{j=1}^L(Z_{1,j}-Z_{2,j})}
\end{align}
that maps a singlet configuration on column $j$ to the un-entangled product state $\vert \up_{1,j}\down_{2,j}\rangle$, and can be implemented via single-qubit rotations followed by two-qubit inter-row interactions. 
After applying $W$, the SWAP operator measurement can then be performed simply by measuring all qubits in the computational basis, and recording $(-1)^{N_{\up\down}}$ where $N_{\up\down}$ is the number of columns measured in the $|\up_{i,1}\down_{i,2}\>$ configuration.

Figure~\ref{fig:timesweeps} numerically shows the expected results of this dynamical string order parameter measurement in the Ising FSPT model, (\ref{eq:3stepdrive}), for various values of $J$ crossing from the FSPT to the trivial, thermal phases. The FSPT regime shows a well-saturated long-time average, with little finite size evolution. By contrast, the trivial and thermal phases show a rapid decay of string-correlations. 

This proposal outlines a practical route towards realizing an FSPT phase and directly measuring the FSPT order that is accessible to the current generation of superconducting qubit devices. The practical limits of topological quantum information storage will require detailed modeling of noise and qubit errors for a given hardware implementation. Noise that is incommensurate with the driving period will tend to melt the MBL bulk and lead to decoherence. In contrast, the topological nature of the FSPT protects entirely against static, symmetry preserving errors, enhancing the fidelity compared to generic quantum memories.

\vspace{2em}\noindent{\it Acknowledgements -- }
We thank C. Neill, P. Roushan, and J. Martinis for insightful conversations.
Numerical simulations were performed at the Texas Advanced Computing Center (TACC) at the University of Texas at Austin. This work was supported by NSF DMR-1653007 (ACP \& AK). This work was performed in part at Aspen Center for Physics, which is supported by National Science Foundation grant PHY-1607611 (ACP \& PTD) and in part at KITP, which is supported by National Science Foundation grant NSF PHY11-25915 (PTD).

\nocite{chen2011complete,turner2011topological,fidkowski2011topological,chen2012symmetry,chen2013symmetry}

\bibliography{FloqSPTbib}

\onecolumngrid
\newpage
\appendix
\twocolumngrid

\section{Generating Ising interactions from two-qubit gates} \label{sec:u1toising}

As discussed in the main text, Floquet SPTs are only possible with discrete symmetries, while the coupling between superconducting qubits using transmon coupling is naturally $U(1)$ symmetric due to the phase symmetry of the superconductor. We can break this $U(1)$ symmetry to $\Z_2$, by applying a series of staggered single-site rotations. Specifically, consider one of the bond application of \eqref{eq:3stepdrive}
\begin{align*}
U_E(J) &=  \exp\left[-i J \sum_{i~\mathrm{even}} \(  X_i  X_{i+1} +Y_i  Y_{i+1}\)\right]. 
\end{align*}
Conjugating this unitary with the staggered $\pi/2$ pulses
\begin{align*}
U_{P} &=   \exp\left[-i \frac{\pi}{2}  \sum_{i~\mathrm{even}} X_i\right].   \nonumber
\end{align*}
changes the relative sign between $XX$ and $YY$ terms in $U_E(J)$. Together with applying $U_E(J')$ with coupling $J'$, this allows independent tuning of the two terms: 
\begin{align*}
&U_{P}^{\dagger} U_{E} (J) U_{P} U_{E}(J') =\\
&  \exp\[-i \sum_{i~\mathrm{even}} \(  \[J + J'\] X_i  X_{i+1} +   \[J - J'\] Y_i  Y_{i+1}  \) \]
\end{align*}
For $J = - J'$ only an $YY$ Ising term remains. In addition to breaking the $U(1)$ symmetry, the sequence of pulses here has the key property that each unitary is still symmetric under the $\Z_2$ symmetry $g = \prod_i X_i$, which protects the FSPT phase. 

\begin{figure}[b]
\centering
\includegraphics[width=0.75\columnwidth]{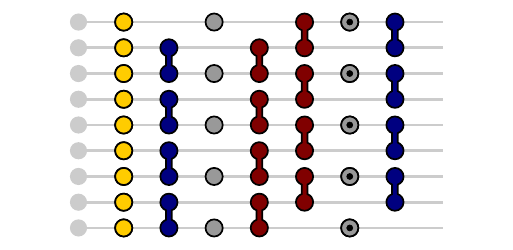}
\caption{{\bf Seven-Pulse Sequence. -- } Sequence implementing a $\Z_2$ FSPT with effective Ising interactions. Time flows left-to-right: $U_D$ (gold), $U_O(-J)$ (blue), $U_P$ (gray), $U_O(J)$ (red), $U_E(J)$ (red), $U_P^\dagger$ (gray with black dot),  $U_E(-J)$ (blue).
 \label{fig:SevenPulsesSeq} 
}
\end{figure}

We can repeat this conjugation with the pulses $U_{P}$ for two-qubit interactions on the odd staggered bonds $U_O(J)$. A modification of the full pulse sequence \eqref{eq:3stepdrive} with only Ising interactions would be
$U_{P}^{\dagger} U_{E} (J) U_{P} U_{E}(-J) U_{P}^{\dagger} U_{O} (J) U_{P} U_{O}(-J) U_{D}$, where $U_D$ are the unitaries with disorder along the $X_i$.  Rearranging this sequences, gives a simpler  equivalent 7-gate sequence, shown in Fig.~\ref{fig:SevenPulsesSeq}.

\section{Dual formulation of the string order parameter \label{app:dual}}
Here, we illustrate the string-order parameter and dual charged-domain wall condensate description for a $1d$ equilibrium (ground-state) SPT with symmetry group: $\Z_n\times \Z_m$, and then construct the analogous properties in a Floquet SPT with symmetry group $\Z_n$. 

\subsection{String order in equilibrium $1d$ SPTs}
We begin by reviewing the classification of $1d$ equilibrium SPTs~\cite{chen2011complete,turner2011topological,fidkowski2011topological} with symmetry group $\Z_n\times \Z_m$, by constructing solvable, fixed point Hamiltonians that realize each of the non-trivial SPT phases in this class based on a decorated domain wall construction~\cite{chen2014symmetry}. These fixed point Hamiltonians enable one to directly deduce the appropriate non-local string order parameters of these phases, and also allow a concrete demonstration that the SPT phases are decorated DW condensates.

Consider a spin chain with $n$- ($m$-) state spins on even (odd) numbered sites respectively. We can describe the $n$-state spins by ``number" operators: $N_{2i}$ with eigenvalues $e^{2\pi ij/n}$ with $j\in \{0,1,\dots n-1\}$, and raising and lowering operators $\eta_{2i}^\pm = \(\eta_{2i}^\mp\)^\dagger$ that increase (decrease) $j$ by $+1\mod n$, i.e. $\eta^-N\eta^+ = e^{2\pi i/n}N$. Similarly define operators $M_{2i+1}$ with eigenvalues $e^{2\pi i k/m}$ with $k\in \{0,1,\dots m-1\}$, and raising and lowering operators $\mu^\pm$ for the $m$-state spins. 

With these ingredients, one can construct a solvable Hamiltonians with SPT ground-states of the form:
\begin{align}
H_{j,k} = -K\sum_{i} \(\mu_{2i-1}^{-k} N_{2i}\mu_{2i+1}^{+k}+\eta^{-j}_{2i} M_{2i+1} \eta^{+j}_{2i+2}\)
\end{align}
where $j\in \{0,1,\dots n-1\}$, and $m\in \{0,1,\dots,m-1\}$.
This Hamiltonian has a $\Z_n\times\Z_m$ symmetry generated by $g_n = \prod_i N_{2i}$, and $g_m = \prod_i M_{2i+1}$, and generalizes the $\Z_2\times\Z_2$ SPT made from spins-1/2: $H_{\Z_2\times\Z_2} = -K\sum_{i} Z_{i-1}X_iZ_{i+1}$.

\subheader{Zero correlation length} One can verify that the Hamiltonian terms commute if $k$ and $j$ are common multiples of $m$ and $n$ and $k = \frac{m}{n}j$. For example: 
\begin{align}
\(\mu_{2i-1}^{-k} N_{2i}\mu_{2i+1}^{+k}\)\(\eta^{-j}_{2i} M_{2i+1} \eta^{+j}_{2i+2}\) = \nonumber\\
e^{2\pi i (k/m-j/n)} \(\eta^{-j}_{2i} M_{2i+1} \eta^{+j}_{2i+2}\)\(\mu_{2i-1}^{-k} N_{2i}\mu_{2i+1}^{+k}\)
\end{align}
in which case we obtain a zero-correlation length Hamiltonian. There exist $\text{LCM}(m,n)$ distinct values of $j$ for which we obtain such commuting Hamiltonians, corresponding to the $\Z_{\text{LCM}(m,n)}$ group structure of SPT phases. To see that these are indeed fixed point Hamiltonians of SPT phases we can examine the structure of edge modes.

\subheader{Edge modes} In a semi-infinite chain with site indices $i=1,2\dots$, there are $\Z_m$ edge mode operators that commute with the Hamiltonian, whose algebra is generated by:
\begin{align}
\tilde{\mu} = \mu_1 ~~~~ \tilde{M}=M_1\eta_2^{+j}
\end{align}
which obey the same algebra as the $\mu$ and $M$ operators of individual $m$-state spins on odd sites. However, unlike the bulk $m$-state spins, these edge modes transform projectively under the $\Z_n$ symmetry. Namely, acting on the edge mode, the $\Z_n$-symmetry generator commutes with $\tilde{M}$ only up to an overall projective phase:
\begin{align}
g_n^{-1} \tilde{M} g_n = e^{2\pi i \frac{j}{n}} \tilde{M} = e^{2\pi i \frac km} \tilde{M} = g_m^{-k} \tilde{M} g_m^{k}
\end{align}
I.e. we can represent the action of the symmetry restricted to the edge modes as: 
\begin{align}
g_{m,\text{edge}} = \tilde \mu ~~~~ g_{n,\text{edge}} = \tilde{M}^k
\end{align}
which form a projective representation of the $\Z_m\times \Z_n$ symmetry:
\begin{align}
g_{m,\text{edge}} g_{n,\text{edge}} = e^{2\pi i \frac{k}{m}} g_{n,\text{edge}}g_{m,\text{edge}} 
\end{align}
Similar considerations show that the right-hand edge of a chain satisfies the conjugate projective representation.

The $\text{LCM}(m,n)$ distinct values of $\frac{k}{m}$ exhaust to the $\text{LCM}(m,n)$ projective representations of the symmetry group $\Z_m\times \Z_n$, which are in one-to-one correspondence with the SPT phases (including the trivial phase $k=0$).

\subheader{Charged domain wall condensation}
The SPT ground-state of $H$ is an eigenstate of each of the operators $\{\mu_{2i-1}^{-k} N_{2i}\mu_{2i+1}^{+k}\}$ with eigenvalue $1$. Hence it is also an eigenstate of product of strings of these operators, such as:
\begin{align}
S^{(m)}_{[i,j]}\equiv \prod_{i<x<j} \mu_{2x-1}^{-k} N_{2x}\mu_{2x+1}^{+k} = \mu_{2i+1}^{-k}\(\prod_{j<i}N_{2j}\)\mu_{2j+1}^{+k}
\end{align}
since $\<S^{(m)}_{[i,j]}\> = +1$ in the ground-state, regardless of $i$ and $j$, the SPT states show long-range string order with respect to $S^{(m)}$.

We next interpret this long range string order as a condensation of domain walls (DWs) of the $\Z_n$ breaking order, bound to $\Z_m$ symmetry charges. Consider taking a $\Z_n$ symmetry-breaking ground-state, $|\text{FM}_a\>$ defined by: $\mu_i|\text{FM}_a\> = e^{2\pi ia/m}|\text{FM}_a\>~\forall i$. There are $n$-distinct such ground-states labeled by $a\in \{0,\dots,n-1\}$, which are related by the $\Z_n$ symmetry. Acting on this state, the string operator $S^{(m)}_I$ creates a DW at the edges of the interval $i$ and also adds charge $\pm k$ to the left and right boundaries respectively. Hence, we can also interpret the long-range order of $S^{(m)}$ as a condensate of these charged DWs.

While we have deduced these properties for the zero-correlation length fixed point Hamiltonian, by definition, an arbitrary FSPT state, $|\psi_\text{SPT}\>$ in the same phase as $H$ will differ only by a finite-depth symmetry preserving local-unitary transformation $U$. I.e for any other SPT ground-state there exist such a $U$ for which: $\<\psi_\text{SPT}|U^\dagger S^{(m)}_{[i,j]}U|\psi_\text{SPT}\> = 1$. Since $U$ commutes locally with the string of symmetry generators $N$ appearing in $S^m$, the transformed string $U^\dagger S^{(m)}_{[i,j]}U$ differs from $S^{m}_{[i,j]}$ only within a region that is exponentially well-localized near the endpoints of the interval $i,j$. Hence, the expectation value of the non-rotated string, will generically be non-zero in any SPT state with the same projective edge symmetry: $\<\psi_\text{SPT}|U^\dagger S^{(m)}_{[i,j]}U|\psi_\text{SPT}\>\geq 0$.

\subsection{Dual formulation of the FSPT string order}
$1d$ FSPTs can also be viewed as equilibrium SPT phases generated by a static Hamiltonian, but with an enlarged symmetry group that incorporates the discrete time-translation symmetry of the drive. This relationship provides a complementary route towards constructing the string order parameter for an FSPT phase. Specifically, the Floquet operator for a $1d$ FSPT, the Floquet operator, $U(T)$, cannot be written as $e^{-iHt}$ for any local, static, symmetry preserving Hamiltonian $H$. However, there generally exists an integer $N$ for which the time-evolution operator for $N$ periods can be written as evolution under a local, symmetric, and time-independent Hamiltonian, i.e.: $\exists ~N~\text{s.t.}~U(NT) = e^{-iNHT}$. However, this effective Hamiltonian, $H$, is not completely arbitrary, but rather, ``remembers" that it comes from time-evolution under $N$ identical driving periods. Namely, in addition to any microscopic symmetry group $G$, $H$ also has an emergent dynamical symmetry generated by: $g= e^{iHT}U(T)$~\cite{yao2017discrete}. This dynamical symmetry satisfies $g^N=1$ (forms a $\Z_N$ group), consists of a product of quasi-local unitary operators due to the MBL nature of $H$ and $U(T)$. Unlike an ordinary microscopic symmetry, however, this dynamical symmetry $g$ is emergent, and its precise form depends on the details of the drive $U(T)$ (it can be explicitly constructed, for example, using a high-frequency expansion). 

The FSPT phases can be understood has SPT phases with this enlarged symmetry group containing both $G$ and $\Z_N$~\cite{else2016classification,potter2016topological}. This relation enables one to classify FSPTs using equilibrium methods such as group-cohomology approaches~\cite{chen2012symmetry,chen2013symmetry}. In addition, this framework provides valuable insight into other non-equilibrium phases, such as discrete time-crystals, which spontaneously break the emergent dynamical symmetry, manifesting in persistent oscillations at a fixed multiple of the fundamental driving period (fraction of the driving frequency).

In fact, the FSPT phases have a dual description in terms of modified time-crystals, which enables a complementary construction of the string order parameter.

Let us consider the dynamical analog of the decorated domain wall construction for a Floquet MBL system. A trivial time-translation invariant Floquet-MBL system (i.e. a time-``liquid" or time-``paramagnet") is a condensate of DWs of time-crystalline order that breaks the emergent time-translation symmetry, i.e. has long-range order in $U_I$. For example, consider a $\Z_2$ time crystal of a spin-1/2 chain, in the limit of zero correlation length, whose time evolution is $U(T)\approx\prod_i X_i e^{-i\sum_i h_iZ_i}$. Starting from a $Z$-basis product state, like $|\cdots\up\up\up\up\up\cdots\>$, the system oscillates with $2T$-periodicity as:
\begin{align}
|\up\up\up\up\up\dots\> \underset{U(T)}{\longrightarrow} |\down\down\down\down\down\dots\>\underset{U(T)}{\longrightarrow}  |\up\up\up\up\up\dots\>
\end{align}
Similarly, starting in the opposite product state, the time-evolution starting from all down exhibits the out-of phase oscillations:
\begin{align}
 |\down\down\down\down\down\dots\>\underset{U(T)}{\longrightarrow}|\up\up\up\up\up\dots\> \underset{U(T)}{\longrightarrow}  |\down\down\down\down\down\dots\>
\end{align}
Then, applying $U_I$ restricted to the middle three spins would alter: $U_I|\up\up\up\up\up\dots\> = |\up\red{\down\down\down}\up\dots\>$. The flipped spins would then oscillate out of phase with the remaining un-flipped spins:
\begin{align}
|\up\red{\down\down\down}\up\dots\> \underset{U(T)}{\longrightarrow} |\down\red{\up\up\up}\down\dots\>\underset{U(T)}{\longrightarrow}  |\up\red{\down\down\down}\up\dots\>
\end{align}
corresponding to an temporally out-of-phase domain sitting inside the larger time-crystal. 

A trivial MBL time-paramagnetic would then be a quantum superposition or ``condensate" of all such domains. where the condensation of time-crystal DWs restores the time-translation symmetry breaking Similarly, an FSPT can be viewed as a condensate of time-crystal domain walls bound to symmetry charges of the microscopic symmetry $G$. The string order parameter $S_I$ is precisely the operator that adds a time-crystal domain in the region $I$ ($U_I$), creating a DW at the right and an anti-DW at the left end, and also inserts a symmetry charge (anti-charge) at the location of the DW (anti-DW).
\end{document}